\documentstyle[newamssymbols]{birk-ot}
\Year{2001}

\def\be{\begin{equation}}
\def\ee{\end{equation}}
\def\ben{\begin{displaymath}}
\def\een{\end{displaymath}}
\def\baa{\begin{eqnarray}}
\def\eaa{\end{eqnarray}}

\def\ba{\begin{array}}
\def\ea{\end{array}}

\def\ep{\varepsilon}
\def\a{\alpha}
\def\g{\gamma}

\def\b{\beta}

\def\e{\varepsilon}

\def\l{\lambda}

\def\th{\vartheta}
\def\Th{\Theta}

\def\O{\Omega}
\def\eb{{\bf e}}
\def\UC{{\hat{T}}}
\def\lp{x}

\def\Sh{\widehat{S}}
\def\Thpq{\Th\left[^\pb_\qb\right]}

\def\B{{\bf B}}
\def\CP1{\C\Bbb{P} 1}

\def\la{\label}

\def\f{\frac}
\def\L{{\cal L}}
\def\p{\partial}
\def\pb{{\bf p}}
\def\qb{{\bf q}}

\def\zb{{\bf z}}
\def\bk{{\bf k}}
\def\tr{{\rm tr}}
\def\0{S}
\def\log{\ln}

\def\w{{\bf w}}
\def\M{{\cal M}}

\def\B{{\bf B}}
\def\la{\label}

\def\f{\frac}
\def\L{{\cal L}}
\def\p{\partial}
\def\res{{\rm res}}

\def\tr{{\rm tr}}
\def\0{S}
\def\1{T}

\def\log{\ln}

\def\bar{\overline}
\def\det{{\rm det}}
\def\dbar{\bar{\partial}}

\begin{document}

\Title{Matrix Riemann-Hilbert problems 
related to branched\\ coverings of $\CP1$}
\Shorttitle{Riemann-Hilbert problems and branched coverings}
\By{{\sc Dmitry Korotkin}}
\Names{Dmitry Korotkin}
\Email{korotkin@mathstat.concordia.ca}
\maketitle

\begin{abstract}
In these notes we solve a class of Riemann-Hilbert (inverse monodromy) problems with an arbitrary quasi-permutation monodromy 
group. The solution is given in terms of  Szeg\"o kernel
on the underlying Riemann surface. In particular, our construction provides a new  class of solutions of the Schlesinger system.
We present some  results on explicit calculation of the corresponding tau-function, and describe divisor of zeros
of the tau-function (so-called Malgrange divisor) in terms of the theta-divisor on the Jacobi manifold of the Riemann 
surface. We discuss the relationship of the tau-function to determinant of Laplacian operator on the Riemann surface.
\end{abstract} 

\newsection{Introduction}

Apart from pure mathematical significance (see review of  {\sc A.A.Bolibruch} \cite{Bolibruch}),
matrix Riemann-Hilbert (inverse monodromy) problems and related theory of isomonodromic deformations
play an important role in  mathematical physics.
In particular, the RH problems 
are central in the theory of integrable systems
(see for example \cite{ZMNP80,Dub94,Hitc94}) and the theory of random matrices \cite{Deift}.
In applications  the main object of interest is the so-called tau-function, which was first introduced 
by {\sc M.Jimbo}, {\sc T.Miwa} and their collaborators \cite{JimMiw81}; it was later shown by {\sc B.Malgrange} \cite{Malg83} that the
tau-function may be interpreted as  determinant of certain T\"oplitz operator. The set of
zeros of the tau-function in the space of singularities of the RH problem is called 
the Malgrange divisor $(\th)$; it plays a crucial role in discussion of solvability
of RH problem with given monodromy data.

For generic monodromy data neither the solution  of a matrix RH problem
nor the corresponding tau-function  can  be computed analytically in terms
of known special functions \cite{Okamoto}. However, there are exceptional cases, when the RH
problem can be solved explicitly; surprisingly enough, these cases often appear in applications.
For example,  the solution of $2\times 2$ RH problem with an arbitrary set of off-diagonal
monodromy matrices was successfully  applied to the problem of finding physically meaningful
solutions of stationary axially symmetric Einstein equations \cite{Koro88,NeuMei,Klein} and to complete classification of 
$SU(2)$-invariant self-dual Einstein manifolds \cite{Hitc94,BabKor98}. The solution of general 
$2\times 2$ RH problem of this kind was given only in 1998 in the papers  \cite{KitKor98,DIKZ98}
(however, some important ingredients of this solution were understood  already  three decades ago \cite{Zver71}). In \cite{KitKor98} it was
also calculated the tau-function corresponding to this RH problem, which turned out to coincide with an  appropriately defined
determinant of Cauchy-Riemann operator acting in certain spinor bundle on underlying  hyperelliptic curve.
In the framework of conformal field theory
this determinant was first introduced  by {\sc Al.B.Zamolodchikov} \cite{Zamo86} (see also \cite{BelKni87,Kniz86,Kniz87,BerRad88,AlMoVa86}).

From the mathematical point of view, determinants of  Cauchy-Riemann operator appear
in the context of holomorphic factorization of determinants of Laplacian operators 
naturally defined via corresponding zeta-function. 
For mathematical description of the determinant bundles over Riemann surfaces  we refer the  reader to  works of {\sc D.Quillen},
{\sc A.A.Beilinson}, {\sc Yu.I.Manin}, {\sc V.V.Schekhtman}, {\sc D.Freed} and other authors (see \cite{Quillen} and references in the memoir by
{\sc J.Fay}  \cite{Fay92}).
In particular, the series of papers  by {\sc L.A.Takhtajan} and {\sc P.G.Zograf} (see lecture notes \cite{TakZog93} and references therein)
contain the most
elementary and simultaneously rigorous treatment of the problem of
holomorphic factorization of Laplacian determinants in the framework of Teichm\"uller theory.

In the  recent paper of the author \cite{Koro00}  the solution of $2\times 2$ case 
 \cite{KitKor98} was generalized  to solve  a class of essentially more non-trivial RH
problems with quasi-permutation monodromies in any matrix dimension. It was also 
conjectured that the coincidence between  corresponding tau-function and determinant of 
appropriately defined
Cauchy-Riemann  operator, observed in the $2\times 2$ case, may be extended to an  arbitrary $N\times N$ case.

Here we give further support to this conjecture, computing the tau-function up to a
nowhere vanishing factor which depends only on the moduli of underlying $N$-fold
covering of the Riemann sphere. 
Comparison with the works  \cite{BelKni87} and \cite{TakZog93} 
suggests   natural interpretation  of  this  factor in the framework of 
the holomorphic factorization of determinants of Laplacians. We would like to notice also  the paper 
\cite{GriOrl89} where the analogy between the  tau-function of Kadomtsev-Petviashvili equation and Cauchy-Riemann determinants
(although rather different from the determinants arising in our context) was observed.

One can hope that in these notes, as well as in the previous works 
\cite{KitKor98,Koro00}, we make a few  steps towards complete solution of one of the problems 
formulated in lecture notes by {\sc V.G.Knizhnik}  \cite{Kniz87} devoted to applications
of geometry of the moduli spaces to perturbative string theory: 
\begin{itemize}
\item
To achieve a complete understanding of the links between 
isomonodromy deformations and  determinants of Cauchy-Riemann operators on  Riemann surfaces.
\end{itemize} 

Let's summarize  some of the results   presented  below in more detail.
Consider an arbitrary compact Riemann surface $\L$ realized as an $N$-sheeted  branched covering of the Riemann sphere.
Denote the coordinate on the Riemann sphere by $\l$ and projections of the branch points on  the Riemann sphere by $\l_1,\dots,\l_M$. 
Then the solution $\Psi(\l)$ of the  inverse monodromy problem with a set of $N\times N$ quasi-permutation\footnote{A matrix is called matrix of quasi-permutation if each of its raws and each of its columns contain only one non-vanishing entry.}
 monodromy matrices, corresponding to the singular points
 $\l_1,\dots,\l_M$, can be written in the following form (this formula is slightly generalized in the main text to allow an arbitrary choice of
non-vanishing entries of the quasi-permutation monodromies):
\ben
\Psi(\l)_{jk}=S(\l^{(j)},\l_0^{(k)})E_0(\l,\l_0)\;,\hskip0.5cm j,k=1,\dots,N\;,
\een 
where $\l^{(j)}$ denotes the point on the $j$th sheet of $\L$, having projection $\l$ on $\CP1$. Here 
$S(P,Q)$ is the  Szeg\"{o} kernel (the reproducing kernel of the $\dbar$ operator acting in a spinor bundle over $\L$):
\ben
S(P,Q) = \f{1}{\Th\left[^\pb_\qb\right](0)}\f{\Th\left[^\pb_\qb\right](U(P)-U(Q))}
{E(P,Q)}\;;
\een
$\Th\left[^\pb_\qb\right](\zb|\B)$  is the theta-function on $\L$  ($\B$ is the matrix of $b$-periods on $\L$) with the argument $\zb\in\C^g$
and characteristics $\pb,\qb\in\C^g$;
$E(P,Q)$  ($P,Q\in\L$) is the prime-form on $\L$ and 
$E_0(\l,\l_0)=(\l-\l_0)/\sqrt{d\l d\l_0}$ is the prime-form on $\CP1$, appropriately lifted to  $\L$. The constant vectors
$\pb,\qb\in\C^g$ (where by $g$ we denote the genus of $\L$) are  such that the combination $\B\pb+\qb$
 does not belong to the theta-divisor $(\Th)$ on the  Jacobi variety
$J(\L)$. 

As follows from the  Fay  identity for the  Szeg\"o kernel \cite{Fay73}, the function $\Psi(\l)$ has determinant $1$ and is normalized at $\l=\l_0$ by the condition
$\Psi(\l=\l_0)=I$. It solves the inverse monodromy problem with quasi-permutation monodromy matrices 
which can be expressed in terms of $\pb,\qb$ and intersection indexes of certain contours on $\L$.
If parameter vectors $\pb$ and $\qb$ (and, therefore, also the monodromy matrices) don't depend on $\{\l_j\}$ then the residues $A_m(\{\l_n\})$ of
the function $\Psi_\l\Psi^{-1}$ at the singular points $\l_m$  satisfy the Schlesinger system.

The   tau-function, corresponding to this solution of the  Schlesinger system,  has the following form:
\be
\tau(\{\l_m\}) = F(\{\l_m\})\Th\left[^\pb_\qb\right](0|\B)\;,
\la{tauint}\ee
where (holomorphic and non-vanishing outside of hyperplanes $\l_m=\l_n$)  function $F$ depends only on the moduli of Riemann surface $\L$ (i.e. points $\{\l_m\}$) and does not depend on the elements of
of monodromy matrices parametrized by vectors $\pb,\qb$.
If all branch points of the Riemann surface $\L$ have multiplicity $1$ (more general surfaces may be obtained from the  surfaces of this class by simple
limiting procedure), the function $F$ is a solution of the following compatible system of equations:
\be
\f{\p F}{\p\l_m} =\f{1}{24} R(\l_m)\;,
\la{Fint}\ee
where $R$ is the  projective connection of $\L$ corresponding to a natural choice of local coordinates on $\L$ in the neighbourhoods of the  points $\l_m$.
Therefore, $F$ is the generating function of the projective connection in our   system of the local coordinates on $\L$.

The compatibility of  equations (\ref{Fint}), which follows from the Schlesinger system, implies the following non-trivial equations for
the values of projective connection at the branch points:
\be
\f{\p R_m}{\p\l_n}=\f{\p R_n}{\p\l_m}\;,
\la{compint}\ee
which were, probably,  unknown before. The equations (\ref{compint}) are  closely related  to the  analogous equations for the accessory
parameters which arise in the problem of uniformization of punctured sphere (see \cite{TakZog93}).

The function $F$ turns out to be non-vanishing in the space of singularities  outside of the hyperplanes $\l_m=\l_n$; therefore, all the zeros of 
the tau-function (\ref{tauint}) come from the zeros of the theta-function. 
This allows to establish the following simple link between the Malgrange divisor $(\vartheta)$  in $\{\l_m\}$-space
and the theta-divisor $(\Th)$ in Jacobi variety $J(\L)$ of the Riemann surface $\L$:
\ben
\{\l_m\}\in (\vartheta)  \;\;  \Leftrightarrow \;\; \B\pb+\qb \in (\Th)\;.
\een

 In the simplest case of  $N=2$ the factor $F$ can also be calculated explicitly (see \cite{KitKor98}) which leads to the following expression for the tau-function:
\be
\tau(\{\l_m\})=[\det{\cal A}]^{-\f 12}
\prod\limits_{m<n}(\l_m-\l_n)^{-\frac 18}\Theta\left[^\pb_\qb\right](0|\B)\;,
\la{tau0}\ee
where $M=2g+2$; $\l_1,\dots,\l_{2g+2}$ are branch points on the hyperelliptic curve $\L$ defined by equation $w^2=\prod_{m=1}^{2g+2}(\l-\l_m)$;
${\cal A}_{\a\b}=\oint_{a_\a}\f{\l^{\b-1}}{w}$ is a $g\times g$  matrix of $a$-periods of non-normalized
holomorphic differentials on this curve.

According to the general philosophy of holomorphic factorization \cite{Quillen,Kniz87,TakZog93},
in general case the square of the module of  function $F$ should be equal to the determinant of Laplacian operator,
up to the factor $\det\Im\B$ and an appropriate Liouville action. The determinant of Laplacian operator with respect to the  Poincare-Lobachevskii
metric is in turn defined via zeta-function regularization.

The main technical tools used here are kernel functions on Riemann surfaces, Fay identities
and deformation theory of Riemann surfaces. The systematic description of these objects may be found
in Fay's books \cite{Fay73,Fay92}. 

We expect present results to find an application to  the problem of isolating the subclass of physically
reasonable solutions of stationary axially symmetric Einstein-Maxwell system \cite{Koro88} in the
spirit of works \cite{Koro88,NeuMei,Klein}, devoted to vacuum Einstein equations. For Einstein-Maxwell system the 
 matrix dimension of RH problem is equal to  three.
Other potential areas of application are the theory of Frobenius manifolds \cite{Dub94} and random
matrices \cite{Deift}. 

Our results confirm an  existence of  deep internal connection between the  algebro-geometric approach to
integrable systems \cite{BBIEM} and certain aspects of conformal field theory.

Let's say a few word about organization of these notes.
In section 2 we remind the formulation of general Riemann-Hilbert (inverse monodromy problem),
associated isomonodromy deformation equations (Schlesinger system), and definition  of Jimbo-Miwa tau-function.
We further 
discuss  quasi-permutation monodromy representations and their natural
relationship to  branched coverings of $\CP1$.

In section 3  we review basic facts from the deformation theory of Riemann surfaces and
adjust them to the situation when the Riemann surface is realized as a branched covering of the
complex plane. Then the moduli space of the Riemann surfaces (more precisely, corresponding Hurwitz space) 
can be parametrized by the projections of  branch points on $\CP1$.

In section 4 we  solve explicitly a class of RH problems  corresponding to an arbitrary 
quasi-permutation monodromy 
representation such that the  associated branched covering possesses the structure of compact
Riemann surface.

In section 5 we prove formula  (\ref{tauint}) for the tau-function and show that the equations (\ref{Fint}) for function 
$F$ can be integrated for the simple case of $2\times 2$
monodromies to give (\ref{tau0}). Here we also discuss  general case.

\section{Riemann-Hilbert  problem with quasi-permutation monodromies and algebraic curves}

\subsection{Riemann-Hilbert problem, isomonodromy deformations and tau-function}

Consider a set of 
$M+1$ points $\l_0,\l_1,\dots,\l_M\in\C$, and a given $GL(N)$  monodromy representation $\M$ of
$\pi_1[\CP1\setminus\{\l_1,\dots,\l_M\}]$. Let us formulate the following Riemann-Hilbert problem:

{\it Find function $\Psi(\l)\in GL(N,\C)$, defined on universal cover of $\CP1\setminus\{\l_1,\dots,\l_M\}$,
which satisfies the following conditions: }
\begin{enumerate}
\item
 $\Psi(\l)$ is normalized at point $\l_0$ on some sheet of the universal cover as follows:
\be
\Psi(\l_0)=I\;;
\la{norm}\ee
\item
$\Psi(\l)$  has right holonomy $\M_\g$ along  contour $\g\in\pi_1[\CP1\setminus\{\l_1,\dots,\l_M\}]$ for all $\g$;
\item
$\Psi(\l)$  has regular singularities of the following form at the points $\l_n$:
\be
\Psi(\l)= \{G_n+ O(\l-\l_n)\}(\l-\l_n)^{T_n} C_n\;,\hskip0.7cm \l\sim\l_n\;,
\la{regsing}\ee
where $G_n, C_n\in GL(N,\C)$; $T_n={\rm diag}(t^{(1)}_n,\dots t^{(N)}_n)$. 
\end{enumerate}
\vskip0.3cm
Consider the following set of standard generators $\g_1,\dots,\g_M$ of
$\pi_1[\CP1\setminus\{\l_1,\dots,\l_M\}]$. Choose  $\l_0$ to be starting point and assume that the contour $\g_n$
starts and ends at $\l_0$ such that interior of $\g_n$ contains only one marked point $\l_n$
(our convention is that the point $\l=\infty$ belongs to the exterior of any closed contour on $\CP1$). 
Moreover, we  assume that  these generators are ordered according to the following relation:
\be
\g_M\g_{M-1}\dots \g_1 = {\bf 1}\;.
\la{rel}\ee 
The matrices $\M_{\g_m}\equiv \M_m$ are called monodromy matrices; as a consequence of (\ref{rel})  we have:
\be
\M_{M}\M_{M-1}\dots \M_{1} = I\;.
\la{Moninf}
\ee

Monodromy matrices $\M_{n}$ are related to coefficients of asymptotics (\ref{regsing}) as follows:
\be
\M_{n}= C_n^{-1} e^{2\pi i T_n} C_n\;.
\ee 
The set of the matrices  $\{\M_{n}\;,\; T_n\;,\; n=1,\dots,M\}$ is called the set of monodromy data. 

Solution  $\Psi(\l)$ of this RH problem satisfies the following  matrix differential equation with meromorphic coefficients:
\be
\f{d\Psi}{d\l}=\sum_{n=1}^M \f{A_n}{\l-\l_n}\Psi\;,
\la{eql}\ee
where
\be
A_n= G_n T_n G_n^{-1}\;.
\la{An}\ee
Suppose now that all monodromy matrices don't depend on positions of singularities  $\{\l_n\}$ and that
for any $n$ none of the numbers $t_n^{(j)}$ differ by integer. Then function $\Psi$, in addition to (\ref{eql}),
 satisfies  the equations with respect to positions of singularities $\l_n$:
\be
\f{d\Psi}{d\l_n}=\left(\f{A_n}{\l_0-\l_n}-\f{A_n}{\l-\l_n}\right)\Psi\;.
\la{eqln}\ee
Compatibility conditions of equations (\ref{eql}) and (\ref{eqln}) imply dependence of 
residues $A_n$ on $\{\l_m\}$ described  by the system of Schlesinger equations:
\ben
\f{\p A_n}{\p \l_m}= \f{[A_n,\,A_m]}{\l_n-\l_m} - \f{[A_n,\,A_m]}{\l_0-\l_m}\;,\hskip0.6cm
m\neq n\; ;
\een
\be 
\f{\p A_m}{\p \l_m}= -\sum_{n\neq m}\left(\f{[A_n,\,A_m]}{\l_n-\l_m} - 
\f{[A_n,\,A_m]}{\l_n-\l_0}\right)\;.
\la{Schl}\ee
Once a solution of the Schlesinger system is given, one can define the locally holomorphic tau-function \cite{JimMiw81}
by the system of equations
\be
\f{\p}{\p\l_n}\log\tau = H_n\equiv \f{1}{2}{\rm res}|_{\l=\l_n}\tr\left(\Psi_\l\Psi^{-1}\right)^2\;;
\hskip0.8cm 
\f{\p\tau}{\p\bar{\l_n}} = 0\;.
\la{taudef}\ee

The tau-function does not depend on normalization point $\l_0$. Namely, function $\Psi^*(\l)$,
corresponding to the same monodromy data and  normalized at a different point $\l_0^*$,
has the form $\Psi^*(\l) =\Psi^{-1}(\l_0^*)\Psi(\l)$. Thus $\tr (\Psi^*_\l{\Psi^*}^{-1})^2=
\tr (\Psi_\l\Psi^{-1})^2$.

Another observation which we shall need below is that  tau-functions corresponding to
monodromy data $\{\M_m\;,\; T_m\}$ and $\{\tilde{\M}_m=D\M_m D^{-1}\;,\; T_m\}$,
where $D$ is an arbitrary  non-degenerate matrix, independent of $\l$ and $\{\l_m\}$,
 coincide. Namely, the new set of monodromies corresponds to function $\tilde{\Psi}=\Psi (\l) D$,
whose logarithmic derivative with respect to $\l$ coincides with the logarithmic derivative of $\Psi$.

According to Malgrange \cite{Malg83}, the isomonodromic tau-function can be interpreted as determinant
of certain T\"oplitz operator. 
The important role in the theory of RH problems is played by the divisor of zeros of the tau-function in the universal covering of the space 
$\{\{\l_m\}\in \C^M\,\big|\;
\l_m\neq\l_n\;\;if \;\; m\neq n\}$. In analogy to the  theta-divisor $(\Th)$ on a Jacobi variety, Malgrange denoted this divisor by $(\th)$.
The importance of the Malgrange divisor $(\th)$ follows from the following fact: if  $\{\l_n\}\in (\th)$,
the Riemann-Hilbert problem with the given set of monodromy matrices and eigenvalues $t_n^{(j)}$
does not have a solution. A close link between Malgrange  divisor $(\th)$ and  theta-divisor $(\Th)\in J(\L)$
for the class of quasi-permutation monodromy representations will be established in sect. \ref{sectau}.

\subsection{Quasi-permutation monodromy representations and branched coverings}

In this paper we shall consider  two special kinds of $N\times N$ monodromy representations.
\begin{definition}\la{permut}
Representation $\M$ is called  the permutations representation if matrix $\M_{\g}$ is a  permutation matrix
for each $\g\in\pi_1[\CP1\setminus\{\l_1,\dots,\l_M\}]$. 
\end{definition}
Remind that a matrix is called  matrix of permutation if each raw and each column of this matrix
contain exactly one non-vanishing entry and this entry equals to 1. 
Permutation matrices are in natural  one-to-one correspondence with elements of permutation group $S_N$.

The definition (\ref{permut}) is self-consistent since the
product of any two permutation matrices  is again a permutation matrix. 

Let us introduce now the notion of  quasi-permutation  monodromy representation:
\begin{definition}
Representation $\M$  is called the  quasi-permutations 
representation if $\M_\g$ is a quasi-permutation matrix for any $\g\in\pi_1[\CP1\setminus\{\l_1,\dots,\l_M\}]$.
\end{definition}
Again, this definition is natural since all quasi-permutation matrices form a subgroup in $GL(N,\C)$. 
We repeat once more that a  matrix is called the   quasi-permutation matrix if each raw and each column of this matrix
contain only one non-vanishing entry.

We shall call two quasi-permutation representations $\M$ and $\widehat{\M}$ equivalent if 
there exists some diagonal matrix $D$ such that
\be
\widehat{\M}_{\g}=D \M_{\g} D^{-1}
\la{equiv}\ee
for all $\g\in \pi_1[\CP1\setminus\{\l_1,\dots,\l_M\}]$.

Since $\det D$ cancels out in (\ref{equiv}), the action of matrix $D$ in (\ref{equiv}) depends on $N-1$  constants.
Therefore, taking
(\ref{Moninf}) into account, we conclude that the $GL(N)$ quasi-permutation representations of 
$\pi_1[\CP1\setminus\{\l_1,\dots,\l_M\}]$ form a $MN-2N+1$ - parametric family.

Let us now discuss the
correspondence between the quasi-permutation representations 
of $\pi_1[\CP1\setminus\{\l_1,\dots,\l_M\}]$
and $N$-sheeted coverings of $\CP1$.

Let $\M$ be a quasi-permutation representation of $\pi_1[\CP1\setminus\{\l_1,\dots,\l_M\}]$.
To every such  representation we can naturally assign a permutation representation $\M'$ substituting $1$ instead of all 
non-vanishing entries  of all monodromy matrices. 
Notice that if some monodromy matrix $\M_n$ is diagonal, the corresponding
element $\M_n'$ of the permutation group is identical.

\begin{proposition}\la{repcurve}
There exists a one-to-one correspondence between permutation representations of
$\pi_1[\CP1\setminus\{\l_1,\dots,\l_M\}]$ and $N$-sheeted compact Riemann surfaces realized as  ramified coverings of $\CP1$ with
projections of branch points  on $\CP1$ equal to $\l_1,\dots,\l_M$.  
\end{proposition}
\begin{proof}
Given a ramified covering 
$\L$ with projections $\l_1,\dots,\l_M$ of branch points on $\CP1$, we
construct the corresponding permutation representation as follows. 
Denote the projection 
of $\L$ on $\CP1$ by $\Pi$.
Generators $\M'_n$ of permutation monodromy group are given by the following construction. Consider the pre-image  $\Pi^{-1}(\g_n)$ of the generator
$\g_n$.
This pre-image is a union of $N$ (not necessary closed) disjoint contours on $\L$ which start and end at
some of the
points $\l_0^{(j)}$ (by $\l^{(j)}$ we  denote the point of$j$th sheet of  $\L$ 
which  has projection $\l$ on $\CP1$). Denote by $\g_n^{(j)}$ the component of 
$\l^{-1}(\g_n)$ which starts at the point $\l_0^{(j)}$; the endpoint of this
contour is $\l_0^{(j_n)}$ for some $j_n\equiv j_n[j]$.
If $\l_n^{(j)}$ is not a branch point, then $j_n[j]=j$, and contour $\g_n^{(j)}$ is closed; if $\l_n^{(j)}$ is a branch point, then 
$j_n[j]\neq j$ and contour $\g_n^{(j)}$ is non-closed.
Then the monodromy matrix $\M'_n$ has 
the following form:
\be
\left(\M'_{n}\right)_{jl} =\delta_{j_n[j], l}
\la{Mperm}
\ee
and naturally  corresponds to some element $s_n$ of the permutation group $S_N$.
On the other hand, starting from some permutation monodromy representation we obviously can  
glue the sheets of the Riemann surface at the  branch points $\{\l_n\}$ in such a way that
it corresponds to  the permutation monodromies (\ref{Mperm}). Moreover, this Riemann surface is obviously compact.
\end{proof}

It is also clear that the branched covering $\L$ is connected iff the associated  permutation representation $\M'$ is
irreducible. 
\begin{remark}
We notice that if some quasi-permutation monodromy matrix $\M_n$ is diagonal, then corresponding matrix $\M'_n$ is equal to 
$I$, and $\l_n$ is in fact  not a projection of 
any branch point on $\CP1$.  However, in the sequel we shall treat such points  in the same
fashion as all other $\l_m$'s. Our  formulas below  explicitly contain multiplicities of all branch points; the formulas
 are written in such form that this does not lead to any inconveniences or inconsistencies.
\end{remark}

In this paper we shall make
two non-essential  simplifying assumptions:
\begin{itemize}\rm
\item
First, we assume that different branch points $P_m$  have different projections $\l_m\equiv\Pi(P_m)$ on $\l$-plane, i.e. $\l_m\neq\l_n$ for $m\neq n$.
\item
Second, we assume that  all the  branch points $P_m$ are {\it simple} (i.e. have multiplicity $1$ or, in other words, we assume that only two sheets coalesce
at each $P_m$).
\end{itemize}

On the level of corresponding permutation representation these assumptions  mean that the group element $s_m$ for each $m$
acts as elementary permutation of only two numbers  of the set $(1,\dots,N)$.
An arbitrary RH problem with
 quasi-permutation representation, corresponding to non-singular curves, may be easily solved by 
degeneration of the construction presented below to 
a submanifold in the space of branch points where some of $\l_m$'s coincide. 
This follows, in particular, from possibility to represent any element of the permutation group 
$S_N$ as a product of the elementary permutations.

According to the  Riemann-Hurwitz formula the genus of  the Riemann surface $\L$ is equal to
\be
g=\frac{M}{2} - N + 1\;;
\la{genus}\ee
therefore, our assumptions about the structure of the covering $\L$  imply, in particular, that the number $M$ is even.

\section{Basic objects on Riemann surfaces. Variational formulas}

\subsection{Basic objects}

Here we collect some useful facts from the theory of Riemann surfaces and their 
deformations.
Consider a canonical basis of cycles  $(a_\a,b_\a),\;\a=1,\dots,g$ on $\L$. Introduce the dual basis of 
holomorphic 1-forms $w_\a$ on $\L$
normalized by $\oint_{a_\a}w_\b=\delta_{\a\b}$.  The matrix of $b$-periods $\B$ and the Abel map $U(P)\,,\;P\in\L$
are given by
\be
\B_{\a\b}=\oint_{b_\a} w_{\b}\;,\hskip0.7cm 
U_\a (P)=\int_{P_0}^P w_\a\;,
\la{Bw}\ee
where $P_0$ is a basepoint.
Consider theta-function with characteristics $\Thpq(\zb|\B)$, where $\pb,\qb\in\C^g$ are vectors of 
characteristics; $\zb \in\C^g$ is the argument. The theta-function is holomorphic function of variable $\zb$ with the
following periodicity properties:
\ben
\Thpq(\zb+\eb_\a)=   \Thpq(\zb) e^{2\pi i p_\a} \;;
\een
\be
\Thpq(\zb+\B\eb_\a)=   \Thpq(\zb) e^{-2\pi i q_\a}e^{-2\pi i z_\a-\pi i \B_{\a\a}}\;,
\la{perth}\ee
where $\eb_\a\equiv (0,\dots,1,\dots,0)$ is the standard basis in $\C^g$. 
The theta-function satisfies the heat equation:
\be
\f{\p^2\Thpq(\O)}{\p z_\a\p z_\b}=4\pi i\f{\p\Thpq(\O)}{\p\B_{\a\b}}\;.
\la{heat}\ee
Let us consider some non-singular half-integer characteristic $[\pb^*,\qb^*]$. The prime-form $E(P,Q)$ is the following skew-symmetric
$(-1/2,-1/2)$-form on $\L\times\L$:
\be
E(P,Q)=\f{\Th\left[^{\pb^*}_{\qb^*}\right](U(P)-U(Q))}{h(P) h(Q)}\;,
\la{prime}\ee
where the square of a section $h(P)$ of a spinor bundle over $\L$ is given by the following expression:
\footnote{One can prove that all the zeros of the r.h.s. of (\ref{hp}) are of the second order; this allows to  define consistently its square root.}
\be
h^2(P)=\sum_{\a=1}^g \p_{z_\a}\left\{\Th\left[^{\pb^*}_{\qb^*}\right](0)\right\} w_a(P)\;.
\la{hp}\ee
To completely define $h(P)$ we assume it to be a section of the spinor bundle corresponding to characteristic
$[^{\pb^*}_{\qb^*}]$.  Then automorphy factors of the prime-form along all cycles $a_\a$ are trivial;
the automorphy factor along each cycle $b_\a$ equals to  $\exp\{-\pi i B_{\a\a}- 2\pi i (U_\a(P)-U_\a(Q))\}$. 
The prime-form has the following local behavior as $P\to Q$:
\be
E(P,Q)=\frac{\lp(P)-\lp(Q)}{\sqrt{d\lp(P)}\sqrt{ d\lp(Q)}}(1+ o(1))\;,
\la{asprime}\ee
where $\lp(P)$ is a local parameter.

The meromorphic symmetric bidifferential on $\L\times\L$ with second
order pole at $P=Q$ and biresidue $1$,
given by the formula
\ben
\w(P,Q)=d_P d_Q\log E(P,Q)\;,
\een
 is called the Bergmann kernel. All $a$-periods of $\w(P,Q)$ with respect to any of its two variables vanish.
The period of Bergmann kernel along basic  cycle $b_\a$ with respect to, say, variable $P$, 
is equal to $2\pi i w_\a(Q)$ and vice verse.
 
The Bergmann kernel has double pole with  the following local behavior on the diagonal $P\to Q$:
\be
\w(P,Q)= \left\{\f{1}{(\lp(P)-\lp(Q))^2} + H(\lp(P),\lp(Q))\right\} d\lp(P) d\lp(Q)\;.
\la{defH}
\ee
where $H(\lp(P),\lp(Q))$ is a  non-singular part of $\w$ in each coordinate chart. 

The restriction of function $H$ on the diagonal gives projective connection $R(\lp)$:
\be
R(\lp)= 6 H(\lp(P),\lp(P))\;,
\la{defR}\ee
which non-trivially  depends on the chosen system of local coordinates on $\L$.
Namely, it is easy to verify that the projective connection transforms as follows
with respect to change of local coordinate $\lp\to f(\lp)$:
\be
R(\lp) \to R(f(\lp)) [f'(\lp)]^2 + \{f(\lp),\lp\}
\la{projcon}\ee
where 
\ben
\{f(\lp),\lp\}\equiv \frac{f'''}{f'}-\f{3}{2}\left(\frac{f''}{f'}\right)^2
\een
is the Schwarzian derivative. 

Suppose that the Riemann surface $\L$ is realized as branched covering of $\l$-plane, and the local coordinates are chosen in
standard way, i.e. $\lp=\l- \Pi(P)$ for any point $P$ which does not coincide with branch points, and 
$\lp=(\l-\Pi(P_m))^{1/\bk_m}$ for any branch point of degree $\bk_m$. 
Projective connection corresponding to this choice of local coordinates will be denoted by $R^H(P)$. 
\footnote{Here ``H'' stands for ``Hurwitz''}

The  Szeg\"o kernel $S(P,Q)$ is the $(1/2,1/2)$-form on $\L\times\L$  defined by the formula
\be
S(P,Q) = \f{1}{\Th\left[^\pb_\qb\right](0)}\f{\Th\left[^\pb_\qb\right](U(P)-U(Q))}{E(P,Q)}\;,
\la{szego}\ee
where $\pb,\qb\in\C^g$ are two vectors such that $\Th\left[^\pb_\qb\right](0)\neq 0$.
The Szeg\"o kernel is the  kernel of the integral  operator $\bar{\partial}^{-1}$, where  the operator $\dbar$  acts in the spinor bundle over $\L$ with
the  holonomies $e^{2\pi i p_\a}$ and $e^{-2\pi i q_\a}$ along basic cycles.
The Szeg\"o kernel itself has  holonomies $e^{2\pi i p_\a}$
and $e^{-2\pi i q_\a}$ along cycles $a_\a$ and $b_\a$, respectively, in its first argument and the inverse holonomies in its second argument.

The Szeg\"o kernel is related to the Bergmann kernel as follows (\cite{Fay73}, p.26):
\be
\hskip0.5cm S(P,Q) S(Q,P) = - \w(P,Q)-\sum_{\a,\b=1}^g\p^2_{z_\a z_\b}\{\log\Thpq(0)\}w_\a (P) w_\b(Q)\;.
\la{SzBerg}\ee

For any two sets   $P_1,\dots,P_N$ and $Q_1,\dots,Q_N$ of points on  $\L$
the following Fay identity takes place (see \cite{Fay73}, p.33):
\be
\det\{S(P_j,Q_k)\}
\la{ident}\ee
\ben
= \frac{\Th\left[^\pb_\qb\right]\left(\sum_{j=1}^N (U(P_j)-U(Q_j))\right)}
{\Th\left[^\pb_\qb\right](0)}\frac{\prod_{j<k} E(P_j,P_k) E(Q_k,Q_j)}{\prod_{j,k} E(P_j,Q_k)}\;.
\een
In particular, for $N=2$ this formula is nothing but  the famous Fay trisecant identity.
The proof of (\ref{ident}) is quite simple: one can check this
identity comparing analytical properties of the l.h.s. and the r.h.s. with respect to all variables $P_j$ and $Q_k$ 
using only the basic facts about holonomies and positions of zeros of the prime-form and theta-function.

Below we  study  dependence of these objects on the moduli of the Riemann surface.
These facts will be required later for  calculation of tau-function. 

\subsection{Variational formulas on a Riemann surface}

 If  a Riemann surface is realized as a branched covering of $\CP1$  then the positions of the branch points may be used
as  natural moduli parameters. The Riemann surfaces which can be obtained  from a given Riemann surface
by variation of positions of the branch points  without changing their
ramification type,
span the so-called Hurwitz spaces (see \cite{Dub94,NatTur00}). 
We shall  start from the  well-known variational formulas on an abstract Riemann surface
and then show how these formulas look in the branched coverings realization.
We shall mainly follow \cite{Fay92}.

Consider a  one-parametric family $\L^{\ep}$  of 
Riemann surfaces of genus $g$. It can be described as smooth deformation of the complex structure
on a fixed Riemann surface $\L^{\e}|_{\ep=0}=\L$. If $\lp$ is a local coordinate on $\L$, the local coordinate $\lp^\e$
on $\L^{\e}$  is holomorphic in $\e$:
\be
\lp^{\e}=\lp +\ep q(\lp,\bar{\lp})+ \dots\;\;.
\la{Taylp}\ee
Then the Beltrami differential $\mu$ (which is a $(-1)$-form with respect to $\lp$ and a $1$-form with respect to $\bar{\lp}$), 
corresponding to the infinitesimal deformation of the curve $\L^\e$ at $\e=0$, is given by
\be
\mu(\lp,\bar{\lp})=\partial_{\bar{\lp}} q (\lp,\bar{\lp})d\bar{\lp}/d\lp\;.
\la{Belt}\ee
Let us introduce the following notations for the infinitesimal deformation defined by the Beltrami differential:
\ben
\delta_\mu\equiv\f{\p}{\p\ep}\Big|_{\ep=0}\;,\hskip0.7cm
\bar{\delta}_\mu\equiv\f{\p}{\p\bar{\ep}}\Big|_{\ep=0}\;.
\een

The infinitesimal variation of the basic  holomorphic 1-forms and matrix of b-periods is given by the following
Rauch  formulas (\cite{Fay92}, p.57):
\baa
\hskip0.3cm \delta_\mu w_\a(Q)= \f{1}{2\pi i}\int_{\L} \mu(P) w_\a(P) \w(P,Q)\;,\hskip0.8cm
 \bar{\delta}_\mu w_\a(Q)= 0\;;
\la{Rauch1}\eaa
\be
\delta_\mu\B_{\a\b}=  \int_{\L} \mu w_\a w_\b\;,\hskip0.8cm
\bar{\delta}_\mu\B_{\a\b}= 0\;.
 \la{Rauch2}\ee
Taking into account that the integral  of the Bergmann kernel  $\w(P,Q)$ along cycle $b_\b$ with respect to variable $Q$ is equal to
$2\pi i w_\b(P)$, the formulas (\ref{Rauch2})  immediately follow from (\ref{Rauch1}).

Let us apply these formulas to a Riemann surface $\L$ realized as a
branched  covering of $\CP1$. We can consider projections 
$\l_m$ of branch points on $\CP1$ as coordinates on the Hurwitz space. 
Therefore, we will be interested in derivatives of basic holomorphic differentials 
and matrix of $b$-periods with respect to $\l_m$.

\begin{theorem}
Basic holomorphic differentials and matrix of $b$-periods of an  $N$-fold covering $\L$ of
$\CP1$   satisfy the following equations with respect to the positions of the branch points:
 \be
\p_{\l_m} \{w_\a(P)\} = \res|_{\l=\l_m} \left\{\f{-1}{(d\l)^2}
\sum_{j} w_\a(\l^{(j)}) \w(\l^{(j)}, P)\right\}\;, 
\la{varw}
\ee
\ben
\p_{\bar{\l}_m} \{w_\a(P)\}=0\;;
\een
\be
\p_{\l_m}\{\B_{\a\b}\}=  \res|_{\l=\l_m} \left\{\f{-4\pi i}{(d\l)^2}\sum_{j< k}{w_\a(\l^{(j)})w_\b(\l^{(k)})}\right\}\;, 
\la{varB1}
\ee
\ben
\p_{\bar{\l}_m}\B =0\;.
\een
\end{theorem}
\begin{proof}
This theorem is valid for arbitrary multiplicities of the branch points. Here we check it under assumption 
that
all branch points are simple and have different projections on $\l$-plane. In this case
the local parameter in the neighborhood of a branch point $\l_m$ is equal to $\lp=\sqrt{\l-\l_m}$; this is the only local parameter
which  depends on the position of $\l_m$. The Taylor series (\ref{Taylp}) looks as follows:
\ben
(\l-\l_m-\ep)^{1/2} = (\l-\l_m)^{1/2} - \f{\ep}{2} (\l-\l_m)^{-1/2} + O(\ep^2)\;.
\een
Therefore, the Beltrami differential 
\be
\mu_m(P)=-\f{1}{2}\left(\p_{\bar{\lp}}\left\{\f{1}{\lp}\right\}\right)\f{d\bar{\lp}}{d\lp} \equiv -\f{\pi}{2}\delta(\lp)
\f{d\bar{\lp}}{d\lp}\;,
\la{mum}\ee
where $\delta(\lp)$ is two-dimensional delta-function, describes the infinitesimal deformation of complex structure under variation of position of 
the branch point $\l_m$
\cite{Fay92}:
\be
\delta_{\mu_m} = \p_{\l_m}\;.
\la{muml}\ee
Substitution of Beltrami differential (\ref{mum}) into Rauch variational 
formulas(\ref{Rauch1}), ((\ref{Rauch2}) gives  (\ref{varw}) and the following
formula for variation of matrix of $\B$-periods:
\be
\f{1}{2\pi i}\p_{\l_m}\{\B_{\a\b}\}=  \res|_{\l=\l_m} \left\{\f{1}{(d\l)^2}\sum_{j=1}^N w_\a(\l^{(j)})w_\b(\l^{(j)})\right\}\;.
\la{varB}
\ee
In turn, this formula implies (\ref{varB1}) if we take into account the following lemma \ref{holsum}.
\end{proof}

\begin{lemma} \la{holsum}
An arbitrary holomorphic differential $w(P)$ on a compact Riemann surface $\L$, realized as $N$-fold covering of $\CP1$, 
satisfies the following relation:
\be 
\sum_{j=1}^N w (\l^{(j)}) =0\;.
\la{sumw}\ee
\end{lemma}

\begin{proof}
This lemma is also valid without any restrictions on ramification type at the points $\l_m$;
it is sufficient to check that $\sum_{j=1}^N w_\a (\l^{(j)})$ is a holomorphic differential on $\CP1$.
The only suspicious points are the branch points $P_m$. 
Regularity of this sum at point $P_m$ follows
 from analysis of its Laurant series in the neighborhood of $\l_m$. 
For example, if the branch point $P_m$ is simple, it is sufficient  to observe
that the local parameter $\lp=\sqrt{\l-\l_m}$ has different signs on the sheets
glued at the branch point $\l_m$.
\end{proof}

\section{Solution of Riemann-Hilbert problems with\\ quasi-permutation monodromies and Szeg\"o kernel}

Here we are going to solve a class  of Riemann-Hilbert problems for 
 quasi-permutation monodromy representations $\M$ which  correspond to branch
coverings with simple  branch points. Solution of a  RH problem corresponding to an arbitrary quasi-monodromy representation 
may be obtained as a limiting case of this construction.
As before, denote projections of the branch points of this curve on $\CP1$ by $\l_1,\dots,\l_M$; assume that
all branch points are simple and have different projections on $\CP1$.
Genus $g$  of the Riemann surface $L$ is equal to $M/2-N+1$; therefore, $M$ should be always even.

In the sequel it will be convenient to assign degree to all of the points $\l_m^{(j)}$ in the following way:
$\bk_m^{(j)}=2$ if $\l_m^{(j)}$ is a (simple!) branch point, and $\bk_m^{(j)}=1$ if $\l_m^{(j)}$ is not a branch point.

Let us introduce on $\L$ a contour $S$, which connects certain 
initial point $P_0$ (it is convenient to assume that $\Pi(P_0)=\l_0$) with all  points $\l_m^{(j)}$, including all branch points.
Suppose that the point $\l_0$ does not belong to the set of projections of basic cycles $(a_\a,b_\a)$ on
$\CP1$. Introduce the following objects:
\begin{itemize}
\item
Intersection indexes of  the contours $l_m^{(j)}$ with all the basic cycles and the contour $S$:
\be
I_{m\a}^{(j)}=l_m^{(j)}\circ a_\a\;,\hskip0.6cm
J_{m\a}^{(j)}=l_m^{(j)}\circ b_\a \;,\hskip0.6cm
K_m^{(j)}= l_m^{(j)}\circ S 
\la{inter}\ee
\ben
{\rm where}\;\;\;
m=1,\dots,M\;;\;\;\a=1,\dots,g\;;\;\;j=1,\dots,N
\een
The contour $S$ can always be chosen in such a way that $K_m^{(j)}=1$ if $\l_m^{(j)}$ is not a ramification point; if
$\l_m^{(j)}$ is a  branch point, then either $K_m^{(j)}=1$ or $K_m^{(j)}=0$.

\item
Two vectors $\pb,\qb\in \C^g$.
\item
Constants 
$r_m^{(j)}\in \C$ assigned to each point $\l_m^{(j)}$; we assume that the 
constants $r_m^{(j)}=r_m^{(j')}$ coincide if $\l_m^{(j)}=\l_m^{(j')}$ i.e. if $\l_m^{(j)}$ is
a branch point. We require that
\be
\sum_{m=1}^M\sum_{j=1}^N r_m^{(j)} =0
\la{sumr}\ee
Therefore, among  constants $r_m^{(j)}$ we have 
only $MN-2g-2N+1$ independent parameters naturally assigned to non-coinciding points among
$\l_m^{(j)}$. 
\end{itemize}

Hence, altogether we introduced $MN-2N+1$ independent constants $\pb,\qb$ and $r_m^{(j)}$; as we saw above,
this number exactly equals the number of non-trivial parameters carried by the  non-vanishing entries of  the quasi-permutation 
monodromy matrices of our RH problem.

Now we are in position to define $N\times N$ matrix-valued function $\Psi(\l)$ which will
later turn out to solve a Riemann-Hilbert problem. We  define the germ of function $\Psi(\l)$
in a small neighborhood of the normalization point $\l_0$ by the following formula:
\be
\Psi(\l)_{kj}=\Sh\Big(\l^{(j)},\l_0^{(k)}\Big)E_0(\l,\l_0)\;.
\la{psinew}\ee
Here $\Sh(P,Q)$ is a section of certain spinor bundle on $\L\times\L$, given by the following
formula inside of the fundamental polygon of Riemann surface $\L$:
\be
\Sh(P,Q)\equiv\f{\Th\left[^\pb_\qb\right]\left(U(P)-U(Q)+\O\right)}
{\Th\left[^\pb_\qb\right](\O)E(P,Q)}\prod_{m=1}^M \prod_{l=1}^N \left[\f{E(P,\l_m^{(l)})}
{E(Q,\l_m^{(l)})}\right]^{r_m^{(l)}}\;.
\la{phinew}
\ee
By $E_0$ we denote the prime-form on $\CP1$
\be
E_0(\l,\l_0)=
\frac{\l-\l_0}{\sqrt{d\l d\l_0}}\;,
\la{primeC}\ee
naturally lifted to $\L$ (the precise way to lift $E_0$ from $\CP1$ to $\L$ we shall
discuss below); 
\be
\O \equiv \sum_{m=1}^M\sum_{j=1}^N r_m^{(j)} U(\l_m^{(j)})\;.
\la{Om}\ee
The vector $\O$ does not depend on the choice of initial point of the Abel map due to assumption
(\ref{sumr}). The formula (\ref{psinew}) makes sense 
 if $\Th\left[^\pb_\qb\right](\O)\neq 0$.

To  define the function $\Psi$  completely we need to specify how to lift  the spinor $\sqrt{d\l}$ 
from $\CP1$ to $\L$. Being lifted on $\L$, the $1$-form $d\l$  has simple zeros 
at all the branch points $P_m$. Therefore,  $\sqrt{d\l}$ is not a holomorphic  section of a spinor bundle on $\L$. However, we can define it in such a way that the ratio $h(P)/\sqrt{d\l}$ (where $\l=\Pi(P)$; ${h(P)}$ is the spinor used in definition of the prime-form) has trivial automorphy factors along all basic cycles.
This function  has poles of order $1/2$ at the
branch points $P_m$ and 
holonomies $-1$ along small cycles encircling the branch points $P_m$.

Consider now the ratio of two prime-forms
\be
f(P,Q)\equiv \f{E_0(\l,\mu)}{E(P,Q)}\;,
\la{fPQ}\ee
where $\l=\Pi(P)$, $\mu=\Pi(Q)$.
 Consider holonomies of $f(P,Q)$ along cycles $a_{\a}$ and $b_{\a}$ with respect to, say, variable  $P$.
From the previous discussion we conclude that these holonomies 
are equal to $e^{\pi i p_{\a}^*}$ and $e^{-\pi i q_{\a}^*-2\pi i(U_\a(P)-U_\a(Q))}$, respectively. Notice, that these holonomies do depend on the choice of 
the odd half-integer
characteristic $\left[^{\pb^*}_{\qb^*}\right]$, in contrast to the holonomies of the prime-form $E(P,Q)$
itself! In addition, $f(P,Q)$ has   holonomies $e^{2\pi i (\bk_m-1)}=\pm 1$ along small cycles encircling branch points $P_m$.

The following theorem gives a solution to a class of RH problems with 
quasi-permutation monodromies. This
is  the main result of present section:

\begin{theorem}\la{Main}
Suppose that $\Th\left[^\pb_\qb\right](\O)\neq 0$.
Let us analytically continue function $\Psi(\l)$ (\ref{psinew}) from the neighborhood of the normalization point 
$\l_0$ to the universal covering $\UC$ of $\CP1\setminus\{\l_1,\dots,\l_M\}$.
Then the function $\Psi(\l)$ is non-singular and non-degenerate  on $\UC$. 
It  has regular singularities at the points $\l=\l_m$, satisfies the 
normalization condition $\Psi(\l=\l_0)=I$ and 
solves the Riemann-Hilbert problem with  the following quasi-permutation monodromies:
\be
\left(\M_n\right)_{kl}=\exp\left\{2\pi i \{{\bk}_n^{(k)}[r_n^{(k)}+{1}/{2}] -{1}/{2}\}K_n^{(k)}\right.
\la{Mon}
\ee 
\ben
\left. +\sum_{\a=1}^g \{J_{n \a}^{(k)}(p_\a+p_\a^*)-I_{n\a}^{(k)} (q_\a+q_\a^*)\}\right\}\delta_{j_m[k],l}
\een
where all constants $\pb,\qb$ and $r_n^{(k)}$ were introduced above; $j_m^{(k)}$ stands for the number of  sheet 
where the contour $l_m^{(k)}$ ends.
\end{theorem}

\begin{proof}
Choose in the Fay identity  (\ref{ident}) $P_j\equiv \l^{(j)}$ and $Q_k\equiv \l_0^{(k)}$. Then, taking into 
account the holonomy properties of 
the prime-form and asymptotics (\ref{asprime}), we conclude that
\ben
\det\Psi = \prod_{m=1}^M\prod_{j,k=1}^N \left[\f{E (\l^{(j)},\l_m^{(k)})}
{E(\l_0^{(j)},\l_m^{(k)})}\right]^{r_m^{(k)}}
\een
 which, being considered as function of $\l$, does not vanish outside of the points $\l_j^{(k)}$; thus $\Psi\in GL(N)$ if $\l$ does not coincide with any of $\l_m$.
The normalization condition $\Psi_{jk}\left(\l_0\right)=\delta_{jk}$ is an immediate corollary 
 of the asymptotic expansion of the  prime form (\ref{asprime}).

Expressions  (\ref{Mon}) for the monodromy matrices of function $\Psi$  follow from the simple 
consideration of the components of function $\Psi$. Suppose for a moment that the function $\Sh(P,\l_0^{(k)})E_0(\l,\l_0)$,
defined by (\ref{phinew}), would  be a single-valued function on $\L$ (as function of $P\in\L$). Then all monodromy
matrices would  be  matrices of permutation: the analytical continuation of  the matrix element 
 $\Sh(\l^{(j)},\l_0^{(k)})E_0(\l,\l_0)$ along contour $l_m^{(j)}$  would simply give the
matrix element $\Sh(\l^{(j)},\l_0^{(\tilde{k}_m)})E_0(\l,\l_0)$. However, since in fact the function 
$\Sh(P,\l_0^{(k)})E_0(\l,\l_0)$ gains some non-trivial multipliers from crossing the basic cycles $a_\a$, $b_\a$
and contour $S$, we get in (\ref{Mon}) an additional exponential factor.
Its explicit form is a corollary of  the definition of intersection indexes which enter this expression, and periodicity
properties of the  theta-function and the prime-form.
\end{proof}

\begin{remark}
If we assume that all constants  $r_m^{(j)}$ vanish, the formula (\ref{psinew}) may be nicely rewritten in terms of the Szeg\"{o} kernel (\ref{szego}) as follows:
\be
\Psi(\l)_{kj} = S(\l^{(j)},\l_0^{(k)})E_0(\l,\l_0)
\la{psisz}
\ee
where $E_0(\l,\l_0)=(\l-\l_0)/{\sqrt{d\l} \sqrt{d\l_0}}$ is the prime-form on $\CP1$.
\end{remark}

If we now assume that vectors $\pb$, $\qb$ and constants $r_m^{(j)}$ don't depend on $\{\l_m\}$  then the monodromy matrices $M_j$ also
don't carry any $\{\l_m\}$-dependence and the   isomonodromy deformation equations are satisfied.

\begin{theorem}
Assume that  vectors $\pb$ and $\qb$ and constants $r_m^{(j)}$   don't depend on $\{\l_m\}$. Then the
functions 
\be
A_n(\{\l_m\}) \equiv {\rm res}|_{\l=\l_n} \left\{\Psi_{\l}\Psi^{-1}\right\}\;,
\la{solA}\ee
where  $\Psi(\l)$ is defined in (\ref{psinew}), 
satisfy the Schlesinger system (\ref{Schl}) outside of the hyperplanes $\l_n=\l_m$ and a submanifold of codimension one defined by the condition
\be
\B\pb+\qb+\O\in (\Th)\;,
\ee
 where $(\Th)$ denotes the  theta-divisor on Jacobian $J(\L)$.
\end{theorem}

\begin{remark}
The formula (\ref{psinew}) remains valid for solution of  RH problem with an arbitrary quasi-permutation monodromy representation
corresponding to a non-singular branched covering. In other words,  our assumption of simplicity of all branch points is 
non-essential. The expressions for monodromy matrices (\ref{Mon}) also remain valid if we assume that the  degree  $\bk^{(j)}_m$
stands for  the number of sheets glued at the point $\l^{(j)}_m$.
\end{remark}

\section{Isomonodromic tau-function and Cauchy-Riemann \\ determinants}
\la{sectau}

\subsection{Tau-function and projective connection}

According to the definition of the tau-function (\ref{taudef}), let us start with calculation of expression
$\tr\left(\Psi_\l\Psi^{-1}\right)^2$.
Notice that this object is independent of the choice of
normalization point $\l_0$ [substitution of $\l_0$ by another point $\tilde{\l}_0$ corresponds to the
$\l$-independent ``gauge'' transformation $\Psi(\l)\to \tilde{\Psi}(\l)= \Psi^{-1}(\tilde{\l}_0)\Psi(\l)$].

Let us rewrite once more the formula (\ref{psinew}) for $\Psi_{jk}$:
\be
\Psi_{kj}(\l,\l_0)=\Sh(\l^{(j)},\l_0^{(k)})\f{\l-\l_0}{\sqrt{d\l}\sqrt{d\l_0}}
\ee
where $\Sh(P,Q)$ is given by expression (\ref{phinew}).
 Consider the limit $\l_0\to\l$. In this limit the matrix elements of the function $\Psi$ behave as follows:
\be
\Psi_{kj}(\l,\l_0)= \f{\l_0-\l}{d\l}\Sh(\l^{(j)},\l^{(k)})  + O\{(\l_0-\l)^2\}\;,\hskip0.6cm k\neq j
\ee
\be
\Psi_{jj}(\l,\l_0)= 1 + \f{\l_0-\l}{d\l}\left\{W_1(\l^{(j)}) - W_2(\l^{(j)})\right\}\;,
\ee  
where $W_1(P)$ is a linear combination of the basic   holomorphic 1-forms on $\L$:
\be
W_1(P) = \f{1}{\Thpq(\O)} \sum_{\a=1}^g \p_{z_\a}\{\Thpq(\O)\} w_\a (P)\;,
\la{W1}\ee
and $W_2(P)$ is the following meromorphic 1-form with simple poles at the points $\l_m^{(j)}$ and
the residues $r_m^{(j)}$:
\be
W_2(P)=  \sum_{m=1}^M\sum_{j=1}^N r_m^{(j)}d_P
\log E(P,\l_m^{(j)})\;.
\ee

Taking into account independence of the expression $\tr\left(\Psi_\l\Psi^{-1}\right)^2$  on   position of the normalization point 
$\l_0$, we have
\ben
\tr\left(\Psi_\l\Psi^{-1}\right)^2(d\l)^2=
2\sum_{j<k}\Sh(\l^{(j)},\l^{(k)})\Sh(\l^{(k)},\l^{(j)}) + \sum_{j=1}^N \left(W_1(\l^{(j)})-
W_2(\l^{(j)})\right)^2\;.
\een
To transform this expression  we first notice that (\cite{Fay73}, p.26)
\ben
\Sh(P,Q)\Sh(Q,P)= -\w(P,Q)-\sum_{\a,\b=1}^g\p^2_{z_\a z_\b}\{\log\Thpq(\O)\}w_\a (P) w_\b(Q)\;.
\een
Furthermore, since $W_1(P)$ is a holomorphic 1-form on $\L$, the expression 
$\sum_{j=1}^N W_1(\l^{(j)})$ vanishes identically according to Lemma\ref{holsum}; hence
\ben
\sum_{j=1}^N \{W_1(\l^{(j)})\}^2 
= -2\sum_{\stackrel{j,k=1}{j<k}}^N \sum_{\a,\b=1}^g\p_{z_\a}\{\log\Thpq(\O)\}
\p_{z_\b}\{\log\Thpq(\O)\} w_\a(\l^{(j)})w_\b(\l^{(k)})\;.
\een

Similarly, we can conclude that $\sum_{j=1}^N \{W_2(\l^{(j)})\}^2$ is a meromorphic 2-form on $\CP1$ which has
poles only at the points $\l_m$; calculation of its residues gives
\be
\sum_{j=1}^N \{W_2(\l^{(j)})\}^2=\sum_{m,n=1}^M \f{r_{mn}(d\l)^2}{(\l-\l_n)(\l-\l_m)}\;,
\ee
where
\be
r_{mn}=\sum_{j=1}^N r_m^{(j)} r_n^{(j)}\;.
\ee

Therefore, as the first step of our calculation,  we get the following expression:
\be
\f{1}{2}\tr\left(\Psi_\l\Psi^{-1}\right)^2(d\l)^2=
-\sum_{j<k} \w(\l^{(j)},\l^{(k)})
\la{trace}
\ee
\ben
-\f{1}{\Thpq(\O)}\sum_{j< k}\sum_{\a,\b} \p^2_{z_\a z_\b}\{\Thpq(\O)\} w_\a(\l^{(j)})
w_\b(\l^{(k)})
+
\f{1}{2}\sum_{m,n}\f{r_{mn}(d\l)^2}{(\l-\l_n)(\l-\l_m)}
\een
\ben
-\f{1}{\Thpq(\O)}\sum_{\a}\p_{z_\a}\{\Thpq(\O)\}\sum_{m}\sum_{j} r_m^{(j)}w_\a(\l^{(j)})
d_P \log E(P,\l_m^{(j)})\;.
\een

Let us now analyze the Hamiltonians 
$$
H_m\equiv \frac{1}{2}{\rm res}|_{\l=\l_m} \left\{\tr\left(\Psi_\l\Psi^{-1}\right)^2\right\}\;.
$$
Using the heat equation for theta-function (\ref{heat}), we can represent $H_m$  in the following form:
\be
H_m = -\res|_{\l=\l_m}\left\{\sum_{j<k} \w(\l^{(j)},\l^{(k)}) \right\}+ \f{1}{2}\sum_{n\neq m}
\f{r_{mn}}{\l_m-\l_n}
\la{Hm}\ee
\ben
+\f{1}{\Thpq(\O)}\sum_{\a,\b}\f{\p\Thpq(\O)}{\p\B_{\a\b}} \p_{\l_m}
\{\B_{\a\b}\}+\f{1}{\Thpq(\O)}\sum_{\a}\p_{z_\a} \{\Thpq(\O)\}\p_{\l_m}\{\O_\a\}\;,
\een
or, equivalently,
\ben
H_m= -\res|_{\l=\l_m}\left\{\f{1}{(d\l)^2}\sum_{j<k} \w(\l^{(j)},\l^{(k)})\right\}
+ 
\p_{\l_m}\log\left\{\prod_{l<n}(\l_l-\l_n)^{r_{ln}}\Thpq(\O)\right\}\,.
\een
Therefore, we come to  the following 
\begin{theorem}
The tau-function
corresponding to solution (\ref{solA}) of Schlesinger system, is given by
\be
\tau({\l_n}) = F(\{\l_n\}) \prod_{m,n=1}^M (\l_m-\l_n)^{r_{mn}}\Thpq\left(\O|\B\right)\;,
\la{tau00}\ee
where function $F(\{\l_n\})$ does not depend on constants $\pb,\qb$ and
 $r_n^{(j)}$, and satisfies the following system of compatible equations
\be
\p_{\l_m}\log F = -\res|_{\l=\l_m}\left\{\f{1}{(d\l)^2}\sum_{j\neq k}\frac{\w(\l^{(j)},\l^{(k)})}{(d\l)^2}\right\}\;.
\la{F1}\ee
\end{theorem}

One can check that only the non-singular part of the Bergmann kernel  contributes to the residue in expression (\ref{F1}); therefore,
we can further express $\p_{\l_m}\log F$ in terms of the projective connection $R^H$ (here ``$H$'' stands for ``Hurwitz'') on the Riemann surface $\L$ corresponding to
the natural choice of local coordinates on $\L$.
\begin{lemma}
Function $F(\{\l_n\})$, defined by (\ref{F1}), satisfies the following system of compatible equations:
\be
\p_{\l_m}\log F = \f{1}{24}R(\l_m)\equiv
-\f{1}{12\pi}\int_{\L}  \mu_m R^H (d\lp)^2\;,
\la{wRH}
\ee
where 
$$\mu_m= -\f{\pi}{2}\delta(\lp)\f{d\bar{\lp}}{d\lp}\;\hskip0.4cm {\rm with} \hskip0.4cm
\lp=(\l-\l_m)^{1/2}$$ 
is the  Beltrami differential (\ref{mum}) corresponding to variation of the  branch point $\l_m$;
 $R^H(P)$ is the projective connection corresponding to our  choice of local parameters on $\L$:  $\lp=(\l-\l_m)^{1/2}$ in the neighborhood of a  branch point  $P_m$ (all branch points are simple
according to our assumption); $ \delta(\lp)$  is  two-dimensional delta-function.
\end{lemma}
\begin{proof}
Formula (\ref{F1}) can be rewritten in terms of the non-singular part  of the
Bergmann kernel 
(\ref{defH}), which immediately leads to (\ref{wRH}) taking into account the definition of projective connection $R$.$\Box$
\end{proof}
According to this lemma, the function $F$ plays the role of generating function of the projective connection corresponding to the
natural choice of coordinate system on the branched covering $\L$.

\begin{remark}
In the  case of higher multiplicity of branch points the  formula (\ref{wRH}) suffers only minor modification: instead of value of the projective connection at the
point $\l_m$ this formula  contains an appropriate  derivative of $R$ at this point.
\end{remark}

\begin{remark}\rm
Integrability of equations (\ref{wRH}) for the function $F$ 
follows from integrability of the equations (\ref{taudef}) for the isomonodromic tau-function.
We would like to notice that it is rather non-trivial fact from the point of view of
the theory of Riemann surfaces that equations
\be
\f{\p R^H(\l_m)}{\p\l_n}=
\f{\p R^H(\l_n)}{\p\l_m}
\la{compat}\ee
are always satisfied if the Riemann surface $\L$ has only simple branch points (for higher multiplicities the values of the
projective connection in (\ref{compat})  should be substituted by an appropriate derivatives).  
This fact looks analogous to similar equations for accessory parameters which appear
in the uniformization problem of punctured sphere \cite{TakZog93}.
\end{remark}

It is possible to prove  that the function $F$ does not vanish as long as the Riemann surface $\L$ remains compact. Therefore,
in particular, it does not vanish outside of the hyperplanes $\l_m=\l_n$. This allows to claim that the zeros of the tau-function
(\ref{tau00}) coincide with the  zeros of the theta-function $\Thpq\left(\O|\B\right)$.  Therefore, we come to  the following relationship
between the Malgrange divisor $(\vartheta)$ in the space of singularities (corresponding to quasi-permutation monodromy groups considered here)
and  the theta-divisor in Jacobi manifold of the Riemann surface $\L$:
\begin{theorem}
The set of singularities $\{\l_m\}$ belongs to the Malgrange divisor $(\vartheta)$ iff the vector $\B\pb+\qb+\O$ belongs to the theta-divisor $(\Th)$ 
in  the Jacobi manifold $J(\L)$ of the Riemann surface $\L$.
\end{theorem}  
We  remind that in the  expression $\B\pb+\qb+\O$ the $\{\l_m\}$-dependence is hidden inside the  matrix of $b$-periods and the vector $\O$.

\subsection{Function $F$ and holomorphic factorization of determinant of \\ Laplacian operator}

Let us make a few  comments concerning the link of the function $F$ with the determinant of Laplacian operator in spirit of previous works
\cite{Kniz87,TakZog93}.  Consider, for example, the case $g> 1$. 
Let us  denote by $z=f(P)$ the fuchsian uniformization map of the curve $\L$ to the fundamental domain $H/\Gamma$ of a fuchsian group $\Gamma$ 
(by $z$ we denote  the complex coordinate on the upper
half-plane $H$). \footnote{For $g=0$ and $g=1$ function $z(x)$  maps $\L$ to the Riemann sphere or fundamental parallelogramm, respectively.}  Then we can write down the Poincare metric of gaussian curvature $-1$ on $\L$:
\ben
ds^2=\f{dz d\bar{z}}{(\Im z)^2}\equiv e^{\phi(\lp)}d\lp d\bar{\lp}\;,
\een
where  
\be
\phi(\lp)=\log\f{|z'(\lp)|^2}{[\Im z(\lp)]^2}
\la{phiz}\ee
is a real function
in each coordinate chart (notice that $\phi(\lp)$ does transform under coordinate change i.e. it is not a scalar).
Function $\phi(\lp)$ satisfies in each chart the Liouville equation
\ben
\phi_{\lp\bar{\lp}}=\f{1}{2}e^{\phi}\;,
\een
which formally  provides  the extremals of  the the Liouville  action
\be
S=\int_{\L} \Big(|\phi_\lp|^2 + e^{\phi}\Big)d\lp  d\bar{\lp}\;.
\la{Liouv}\ee
However,  since the function $\phi$, defined by (\ref{phiz}), does not behave like  a scalar with respect to the coordinate change, 
this expression has to be accurately defined in each case taking into account the
terms coming from the boundaries of the coordinate charts \cite{TakZog93}. This can, for example, be explicitly done if the local coordinate $\lp$
corresponds to the Schottky uniformization of the Riemann surface  $\L$ \cite{TakZog87}. It is also easy to write down these boundary terms
in system of local parameters associated to  the ramified covering realization, but we will not discuss them here.

The Laplace operator on $H$ 
\ben
\Delta=(z-\bar{z})^2\f{\p^2}{\p z\p\bar{z}}
\een
is invariant with respect to the M\"obius group. Therefore, it can be naturally defined on $\L$, where it is self-adjoint, non-negative,
and has discrete spectrum in the Hilbert space of functions on $\L$. The Hilbert space is equipped with the  natural inner product privided by the Poincare
 metric.
The zeta-function of the Laplacian $\Delta$ is defined in terms of its eigenvalues $\kappa_j$  as follows: $\zeta(s)=\sum_j\kappa_j^{-s}$.
In turn, the  determinant of the Laplacian $\Delta$ is defined via  analytical continuation of the zeta-function to zero:
\ben
\det\Delta \equiv \exp\{\zeta'(0)\}\;.
\een
An infinitesimal variation of the moduli of the curve $\L$ by a Beltrami differential $\mu$ leads to the  following variation of  the determinant of Laplacian operator
\cite{ZogTak87a}:
\be
\delta_\mu \log\f{\det\Delta}{\det\Im\B} =
\f{1}{6\pi}\int_{H/\Gamma} \mu(z) R^F(z)(dz)^2\;,
\la{vardet1}\ee
where $R^F(z)$, $z\in H/\Gamma$ is the  projective connection corresponding to Fuchsian uniformization of $\L$.

On the other hand, the  action of the Beltrami differential $\mu_m$   on the logarithm of the real 
function $|F|^2$ follows from (\ref{wRH}):
\be
\p_{\l_m}\log |F|^2 = 
-\f{1}{12\pi}\int_{\L} \mu_m(\lp) R^H(\lp) (d\lp)^2 \;,
\la{varF}\ee
where $R^H(\lp)$ denotes the projective connection with respect to the natural system of local parameters on $\L$ arising from the  realization of $\L$ as
branched covering of $\CP1$. Therefore, there must exist a  real-valued function $S^H(\{\l_m\})$  such that
\be
|F|^2 = \left(\f{\det\Delta}{\det\Im\B}\exp\left\{\f{1}{12}S^H\right\}\right)^{-1/2}\;.
\la{F2}\ee
 According to (\ref{vardet1}) and (\ref{varF}), the  function $S^H$ satisfies the following compatible system of equations:
\be
\f{\p S^H}{\p\l_m} = 2\int_{\L} \left[R^H(\lp)-R^F(z(\lp)) \Big|\f{d z}{d\lp}\Big|^2\right]\mu_m (d\lp)^2 \;. 
\la{SH1}\ee
Taking into account that the Beltrami differential  $\mu_m$ is, up to the factor  $-\pi/2$,  nothing but
 the  delta-function with support  at the branch point $\l_m$, and 
that the projective connection transforms according to (\ref{projcon}) under a change of the local coordinate, we get the following equations for $S^H$ in terms
of Schwartzian derivative of the local parameters:
\be
\f{\p S^H}{\p\l_m}=  -\{z(\lp_m),\lp_m\}\Big|_{\lp_m=0}\;,
\la{SH2}
\ee
where $\lp_m=\sqrt{\l-\l_m}$ is the local prameter near the branch point.
According to the privious experience (see \cite{TakZog93}, where the relationship between Fuchsian and Shottky uniformizations is discussed in detail,
and more recent paper \cite{AndTak97}), 
the solution of equations  (\ref{SH2}) should coincide   with an appropriately defined
 Liouville action (\ref{Liouv}). In addition to the bulk term (\ref{Liouv}), the Liouville action contains suitable boundary terms, which we don't write down explicitly.

According to the general philosophy of holomorphic factorization \cite{Quillen},  the formula (\ref{F2})
allows to identify $F$ with $[\det\dbar_0]^{-1/2}$. The operator $\dbar_0$ acts on sections of the trivial bundle over $\L$; 
this operator should be understood as differentiation  with respect to our standard  system
of local parameters on the branched covering.

We would like to refer also to work \cite{AndTak97}, where the generating function for projective connection was
 computed with respect to the Fuchsian uniformization; this gives further support to the hypothesis of  close relationship between 
function  $F$ and appropriate version of
Liouville action in our parametrization.

\subsection{Hyperelliptic curves and $2\times 2$ Riemann-Hilbert problems with off-diagonal monodromies }

Here we consider the simplest
case of $N=2$, when any matrix of quasi-permutation is either diagonal or off-diagonal. We shall consider  monodromy groups 
containing only  off-diagonal monodromies; the insertion of additional diagonal monodromies according to the general scheme is straightforward.
In this case the branched covering $\L$ corresponds to hyperelliptic algebraic curve with branch points $\l_1,\dots,\l_M$ and function $F$ 
may be calculated explicitly \cite{KitKor98}.
We have  $M=2g+2$, where  $g$ is the genus of the branched covering $\L$; this branched covering is the Riemann surface of the algebraic curve
\be
w^2=\prod_{m=1}^{2g+2}(\l-\l_m)
\la{he}\ee

It is convenient to put all $r_m^{(j)}=0$;  in this case the  formula (\ref{psisz}) gives the  solution $\Psi(\l)\in SL(2)$ of the RH problem with
arbitrary off-diagonal monodromies having  unit determinant:

\ben
M_m=\left(\ba{cc}                0        &  d_m      \\
                        -d_m^{-1}    &     0        \ea\right) \;,
\een
where constants $d_m$ may be expressed in terms of the  elements of vectors $\pb,\qb$. Let us count the number of essential parameters in
the monodromy matrices and in the construction of function $\Psi$. The matrices $M_m$ contain altogether $2g+2$ constants; however,
there is one relation (product of all monodromies gives $I$). One more parameter is non-essential due to possibility of simultaneous
conjugation of all monodromies with an arbitrary  diagonal constant matrix.
Therefore, the set of monodromy matrices contains $2g$ non-trivial constants in accordance with number of non-trivial constants contained 
in vectors $\pb$ and $\qb$.

To integrate the remaining equations
\be
\p_{\l_m}\log F = \f{1}{24}R^H(\l_m)
\la{FHE}
\ee
 on  hyperelliptic curve (\ref{he}) we use the following formula (\cite{Fay73}, p.20) 
for the  projective connection at arbitrary point of  $P$ of the
hyperelliptic curve $\L$  (where $\lp$ is the local parameter in the neighborhood of the point $P$, $\l=\Pi(P)$ is the projection of $P$ on $\l$-plane):
\be
R^H(P)=\{\l(\lp),\lp\}(P)+\f{3}{8}\left(\f{d}{d \lp}\log\f{\prod_{\l_m\in T}(\l-\l_m)}{\prod_{\l_m\not\in T}(\l-\l_m)}\right)^2 (P)
\la{prhyp}\ee
\ben
-\f{6}{\Th\left[^{\pb^T}_{\qb^T}\right](0)}\sum_{\a,\b=1}^g \p^2_{z_\a z_\b}\left(\Th\left[^{\pb^T}_{\qb^T}\right](0)\right)
\f{w_\a}{d \lp}(P)\f{w_\b}{d \lp}(P)\;.
\een
Here $\{\l,\lp\}$ is the Schwarzian derivative of $\l$ with respect to $\lp$; 
$T$ is an arbitrary divisor consisting of  $g+1$ branch points, which satisfies certain non-degeneracy condition. Characteristic
$\left[^{\pb^T}_{\qb^T}\right]$ is the even half-integer  characteristic corresponding to the divisor
$T$  according to  the following equation:
\be
\B\pb^T+\qb^T=\sum_{\l_m\in T} U(\l_m)-K\;,
\la{pqT}\ee
where $K$ is the  vector of Riemann constants; the initial point of the Abel map is chosen to be, say, $\l_1$. In this case the
r.h.s. of (\ref{pqT}) is a linear combination, with integer or half-integer coefficients, of the  vectors $\eb_\a$ and $\B\eb_a$. These  coefficients
are composed in vectors $\pb_T$ and $\qb_T$. The non-degeneracy  requirement imposed on the divisor $T$ gives rise to the condition that
the vector  $\B\pb^T+\qb^T$
does not belong to the  theta-divisor on $J(\L)$, i.e.
 $\Th\left[^{\pb^T}_{\qb^T}\right](0)\neq 0$.

Of course, the projective connection $R$, as well as the function $F$,  are independent of the  choice of the divisor $T$, which plays only intermediate role.
If in (\ref{prhyp}) we choose $P=\l_m$, the local parameter is $\lp=\sqrt{\l-\l_m}$. Then all terms in $R(\l_m)$
which don't contain theta-function can be integrated explicitly; the terms
 containing theta-function
can be represented as logarithmic derivative with respect to $\l_m$ by making  use of  the heat equation for 
theta-function (\ref{heat}) and Rauch formula (\ref{varB1}). These terms are equal to 
\ben
-6 \f{\p}{\p\l_m}\log\Th\left[^{\pb^T}_{\qb^T}\right](0)\;.
\een
This expression may be rewritten using the Thomae formula \cite{Mumford}
\ben
\Th\left[^{\pb^T}_{\qb^T}\right](0)=\pm(\det{\cal A})^2 \prod_{\l_m,\l_n\in T} (\l_{m}-\l_{n})
 \prod_{\l_m,\l_n\not\in T}(\l_{m}-\l_{n}),
\een
where ${\cal A}_{\a_\b}=\oint_{a_\a}\f{\l^{\b-1}}{w}$ is the $g\times g$ matrix of $a$-periods of non-normalized holomorphic differentials on $\L$.

Collecting together all the explicit factors arising from the Thomae formula and expression (\ref{prhyp}), we get the following answer for the
function $F$:
\be
F=[\det{\cal A}]^{-\f 12}
\prod\limits_{m<n}(\l_m-\l_n)^{-\frac 18}
\la{Fhe}
\ee
which coincides with the expression for the determinant $\{\det\dbar_0\}^{-1/2}$  of Cauchy-Riemann operator acting in trivial bundle over $\L$
with respect to our system of local parameters \cite{Zamo86,Kniz87}. For the tau-function itself we get the following expression
\ben
\tau(\{\l_m\})=[\det{\cal A}]^{-\f 12}
\prod\limits_{m<n}(\l_m-\l_n)^{-\frac 18}\Theta\left[^\pb_\qb\right](0|\B)\;,
\een
which, according to the same papers, coincides with naturally defined determinant of the Cauchy-Riemann operator acting on spinors which have
holonomies $e^{2\pi i p_\a}$ and $e^{-2\pi i q_\a}$ along basic cycles of $\L$.

As we saw above, for general curves the interpretation of the factor $F$ as $\{\det\dbar_0\}^{-1/2}$ can, probably, be preserved. 
Interpretation of the whole tau-function as  $\det\dbar_{1/2}$ in a twisted spinor bundle  remains valid for arbitrary curves, 
if all constants $r_m^{(j)}$ vanish.

\begin{remark}
In the article  \cite{Palmer} it was argued that the tau-function for isomonodromy deformations with arbitrary (not only quasi-permutation)
monodromy matrices can be interpreted as determinant of Cauchy-Riemann operator in appropriate spinor vector bundle over punctured sphere with cuts.
We don't know at the moment how
to establish an explicit link between the framework of \cite{Palmer} and our present scheme, where  the spinor line bundles
over compact Riemann surfaces appear.
\end{remark}

\begin{acknowledgements}
I would like to thank A.Bobenko, J.Harnad, J.Hurtubise, A.Kokotov, V.B.Matveev, A.Orlov and A.N.Tyurin for comments and discussions. 
This work was supported by  NSERC and FCAR grants, and  Laboratoire Gevrey de  Math\'ematique Physique, Universit\'e de Bourgogne. 
I thank V.B.Matveev for hospitality at Universit\'e de Bourgogne, where this work was completed.
\end{acknowledgements}


\address{Department of Mathematics and Statistics\\
Concordia University\\
7141 Sherbrook West, Montreal\\
H4B 1R6 Quebec\\ Canada}


\received{   }

\end{document}